\documentclass[aps, showpacs,preprintnumbers,amsmath,amssymb,superscriptaddress,twocolumn,nofootinbib]{revtex4-2}
\usepackage{amsmath,amsfonts,amssymb,graphics,graphicx,epsfig,color,times,indentfirst,layout,hyperref,multirow,bm}
\usepackage{hyperref, soul}
\hypersetup{linktocpage,colorlinks=true}
\usepackage{threeparttable}
\usepackage{makecell}
\usepackage{tabularx} % Required for tabularx environment
\usepackage{booktabs} % Required for \toprule, \midrule, \bottomrule
\usepackage{ragged2e}
\newcolumntype{L}{>{\RaggedRight\arraybackslash}X}   % Left-aligned X column
\newcolumntype{C}{>{\Centering\arraybackslash}X}     % Centered X column
\newcolumntype{R}{>{\RaggedLeft\arraybackslash}X}    % Right-aligned X column

\usepackage{dcolumn}
\usepackage{bm}
\usepackage{color}
\usepackage{footnotebackref}
\begin{document}
\title{Bulk photovoltaic effects in the Haldane model}% two-dimensional mirror-time symmetric systems}

\author{Bo-Xin Lin}
\affiliation{Graduate Institute of Applied Physics, National Chengchi University, Taipei 11605, Taiwan}

\author{Hsiu-Chuan Hsu}
\email{hcjhsu@nccu.edu.tw}\affiliation{Graduate Institute of Applied Physics, National Chengchi University, Taipei 11605, Taiwan}
\affiliation{Department of Computer Science, National Chengchi University, Taipei 11605, Taiwan}

\date{\today}

\begin{abstract}
	The bulk photovoltaic effect (BPVE) refers to the direct current generation in a noncentrosymmetric material under illumination and can be applied to solar energy technology. BPVE includes injection and shift currents, led by the change of velocity and displacement of wave packet during optical transitions, respectively. We derive the constraints on the conductivity tensors imposed by mirror-time ($\mathcal{MT}$) symmetry for two-dimensional systems. For the Haldane model, we show that linearly polarized light can induce shift and injection currents. In contrast, circularly polarized light can not induce shift or injection currents, as constrained by the three-fold rotation symmetry. Additionally, due to the presence of $\mathcal{MT}$ symmetry, a separation of responses is shown in the Haldane model. Under linearly polarized light, shift current, allowed by time-reversal symmetry, flows perpendicularly to the injection current, allowed by $\mathcal{MT}$ symmetry. Across the topological phase transition, the injection current does not change sign since the group velocity's sign remains unchanged.  On the contrary, shift current shows a sign flip, as a result of band inversion. Furthermore, we calculate quantum geometry, including quantum metric and symplectic connection, to demonstrate the microscopic quantum origin of the BPVE. We found that the vector field of symplectic connection in the Brillouin zone possesses vortices in the topological phase, but not in the trivial phase.  
	%The bulk photovoltaic effect (BPVE) refers to the direct current generation in a noncentrosymmetric material under illumination and can be applied to solar energy technology. BPVE includes injection and shift currents, led by the change of velocity and displacement of wave packet during optical transitions, respectively. We derive the constraints on the conductivity tensors imposed by mirror-time ($\mathcal{MT}$) symmetry for two-dimensional systems. For the Haldane model, we show that linearly polarized light can induce shift and injection currents, which vanish under circularly polarized light as constrained by the three-fold rotation and $\mathcal{MT}$ symmetry. Additionally, due to the presence of $\mathcal{MT}$ symmetry, a separation of responses is observed in the Haldane model: one direction exhibits a time-reversal symmetry-allowed photocurrent, whereas another manifests a parity-time symmetry-allowed photocurrent. Across the topological phase transition, the injection current does not change sign, whereas shift current shows a sign flip. The vector field of the Hermitian connection in the Brillouin zone possesses vortices in the topological phase, but not in the trivial phase.  Furthermore, we calculate the related quantum geometry, including Berry curvature, quantum metric and Hermitian connection, and demonstrate the microscopic quantum origin of the BPVE.  
\end{abstract}
 \maketitle
\section{introduction}
% quantum geometrical origin in solids: from Hall effects to optical responses
In noncentrosymmetric solids, d.c. current can be generated without external bias under light, known as  Bulk photovoltaic effects (BPVE) \cite{Dai2023}. From second-order perturbation theory with length gauge, one can deduce two contributions, injection and shift current, according to the physical interpretation of the formula \cite{Aversa1995,Sipe2000,Dai2023}. The injection current is given by the change of the group velocity during optical transitions between bands, and the shift current is given by the displacement of the Bloch wave packet during the optical transition. In contrast to group velocity, which can be straightforwardly obtained by the slope of the energy bands, the displacement of the wave packet is a property of Bloch wave functions that can not be inferred solely by energy bands. Furthermore, quantum geometry \cite{Morimoto2016a,Ahn2020,Ma2021,Ahn2022,Bhalla2022,Hsu2023,Morimoto2023, Bhalla2023,Jiang2025}, %which play an important role in nonlinear optics, for example, high-harmonic generations \cite{Qian2022} and second-order photovoltaics \cite{Tan2016,Yang2024,Bai2024}.
%Recently, the quantum geometric origins of injection and shift current have been investigated \cite{Ahn2020,Ahn2022,Hsu2023}. Quantum geometry
provides an intrinsic characterization of response functions in solids and elucidates a more complete semiclassical theory for wave packet dynamics \cite{Xiao2010,Gao2014,Smith2022,Jia2024,Yoshida2025Emergent}. It has been shown that the injection current is relevant to the quantum geometric tensor and the shift current to the Riemmanian connections \cite{Ahn2022,Jiang2025}. 

Crystal symmetry analysis is a useful method for determining the nonzero components of the second-order photoconductivity tensors \cite{Boyd,Jiang2025}. For decades, it has been known that in time reversal (TR) symmetric systems, the shift current is driven by linearly polarized light, dubbed linear shift current, and the injection is driven by circularly polarized light, dubbed circular injection current. Nonetheless, it has been shown recently that in parity-time ($\mathcal{PT}$) symmetric systems, the linear shift and circular injection currents are suppressed by $\mathcal{PT}$ symmetry constraint \cite{Ahn2020,Watanabe2021,Ezawa2025}. Other responses are allowed: the shift current is driven by circularly polarized light, called circular shift current, and the injection is driven by linearly polarized light, called linear injection current. Furthermore, in mirror time ($\mathcal{MT}$)-symmetric systems, certain components reflect TR symmetry and others manifest $\mathcal{MT}$ symmetry, called separation of responses \cite{Ahn2020}. In magnetic systems, $\mathcal{MT}$ symmetry could still be preserved even without $\mathcal{PT}$ symmetry, such that the separation of responses is expected to occur. 

% why the importance ?
% GAP:
%Shift in mirror nonmagnetic symmetric systems \cite{Azpiroz2020}. 
There is a great interest in the study of BPVE in two-dimensional systems. A high-throughput numerical calculation of the shift currents for nonmagnetic films in C2DB database \cite{Haastrup2018,Gjerding2021} has been carried out \cite{Sauer2023}. BPVE and the relevant intrinsic quantum geometry in twisted bilayer graphene have been investigated by several groups \cite{Arora2021,Chaudhary2022,Kaplan2022,Penaranda2024}. Because of the sensitivity of the second-order photoconductivity tensors to the underlying symmetry, BPVE can be used as a probe of the symmetry breaking effects, such as strain and strong correlations.  Additionally, BPVE in magnetic materials has gained significant attention, for instance, ferromagnetic topological insulators \cite{Ogawa2016,Bhalla2020}, ferroelectric films \cite{Li2021}, $\mathcal{PT}$ symmetric anti-ferromagnetic CuMnAs \cite{Bhalla2023}, $\mathcal{PT}$ symmetric CrI$_3$ \cite{Zhang2019switchable} and altermagnets \cite{Ezawa2025}.  

As a seminal model for Chern and topological insulators, the Haldane model can be applied to understand essential physics in two-dimensional systems. Theoretical studies with the Haldane model provide profound insight to the phenomena being investigated. The feasibility of experimental realization is particularly high in systems utilizing ultracold fermions \cite{Jotzu2014}, offering a suitable platform for verifying the theoretical predictions in this model.
The shift current in the Haldane model has been theoretically studied \cite{Sivianes2023,Yoshida2025Diverging}. It was found that the shift current changes sign across the topological phase transition \cite{Sivianes2023}. Furthermore, diverging shift current, proportional to the inverse of photon frequency,  is found in the gapless limit in the Haldane model \cite{Yoshida2025Diverging}. However, the generation of injection current, a detailed symmetry analysis and the underlying quantum geometry of the injection and shift conductivity tensors for the Haldane model are lacking. 
 
 %The responses in $\mathcal{MT}$ symmetric 2D system is less systematically discussed. 
 
\begin{figure}
	\includegraphics[width=0.48\textwidth]{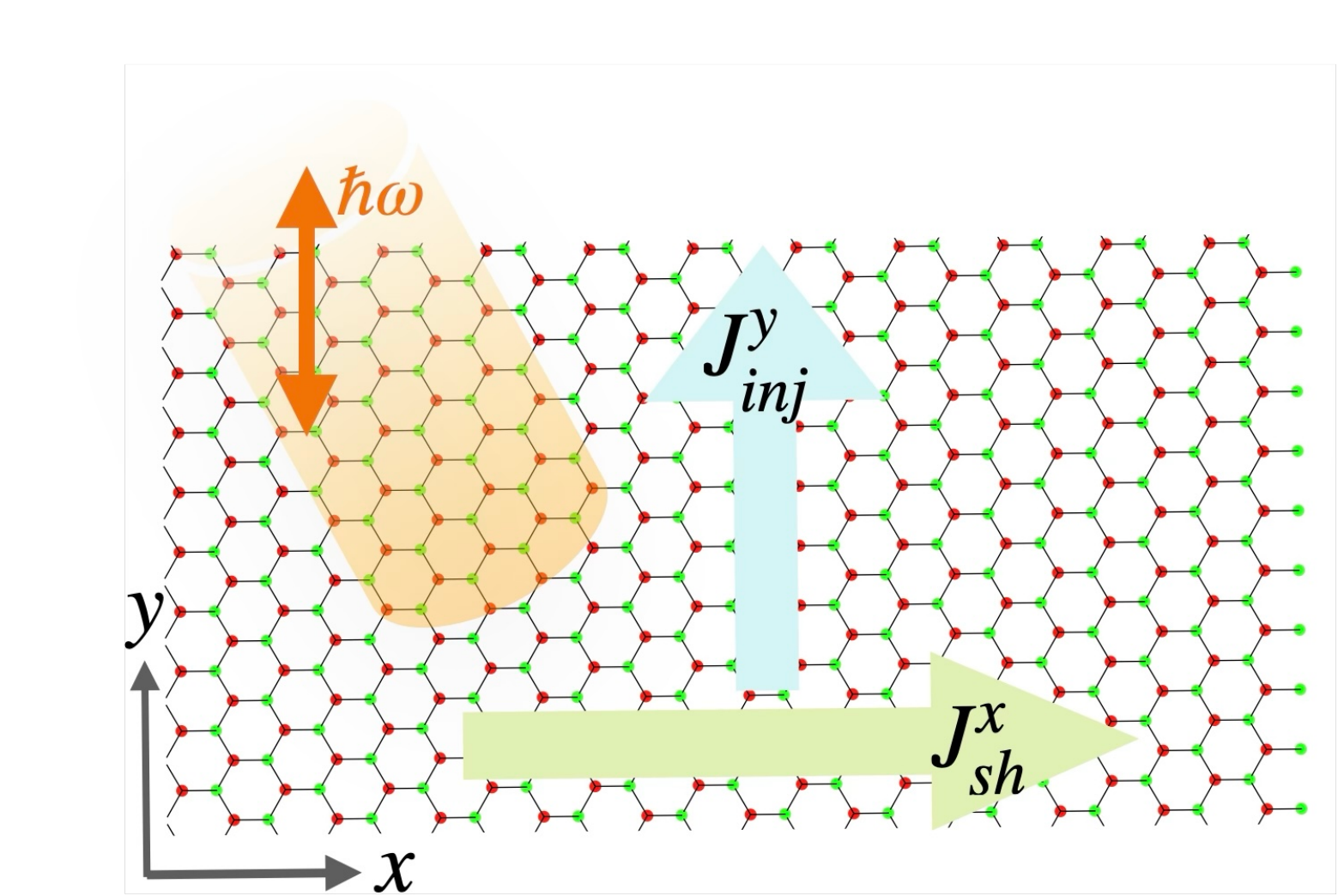}
	\caption{Schematic of response separation in the Haldane model. For a $y$-polarized plane electromagnetic wave incident normally upon the $x$-$y$ plane, the $\mathcal{T}$-allowed shift current flows along $x$-direction, and the $\mathcal{M}_y\mathcal{T}$-allowed injection current flows along $y$-direction.}
	\label{fig:setup}
	\end{figure}

In this study, we discuss the effect of $\mathcal{MT}$ symmetry on injection and shift currents in two-dimensional systems, using the Haldane model as an example. We found that under linearly polarized light, these two currents are orthogonal. The shift current, allowed by $\mathcal{T}$ symmetry, flows perpendicularly to the injection current, allowed by $\mathcal{MT}$ symmetry. A schematic plot of the BPVE in the Haldane model is shown in Fig. \ref{fig:setup}. Furthermore, we carry out numerical calculations of BPVE in the Haldane model. In addition, quantum geometry, including quantum metric and symplectic connections are shown. 
The remainder of this paper is as follows. In Sec. \ref{sec:model}, the Haldane model and the formula for BPVE are introduced. The symmetry analysis on the conductivity tensors of $\mathcal{MT}$ symmetry is given. In Sec. \ref{sec:results}, the numerical results are presented. Lastly, a conclusion is given in Sec. \ref{sec:disc}.

\section{Model and method}\label{sec:model}
\subsection{The Haldane model}\label{sec:hamiltonian}
\begin{figure}
	\includegraphics[width=0.48\textwidth]{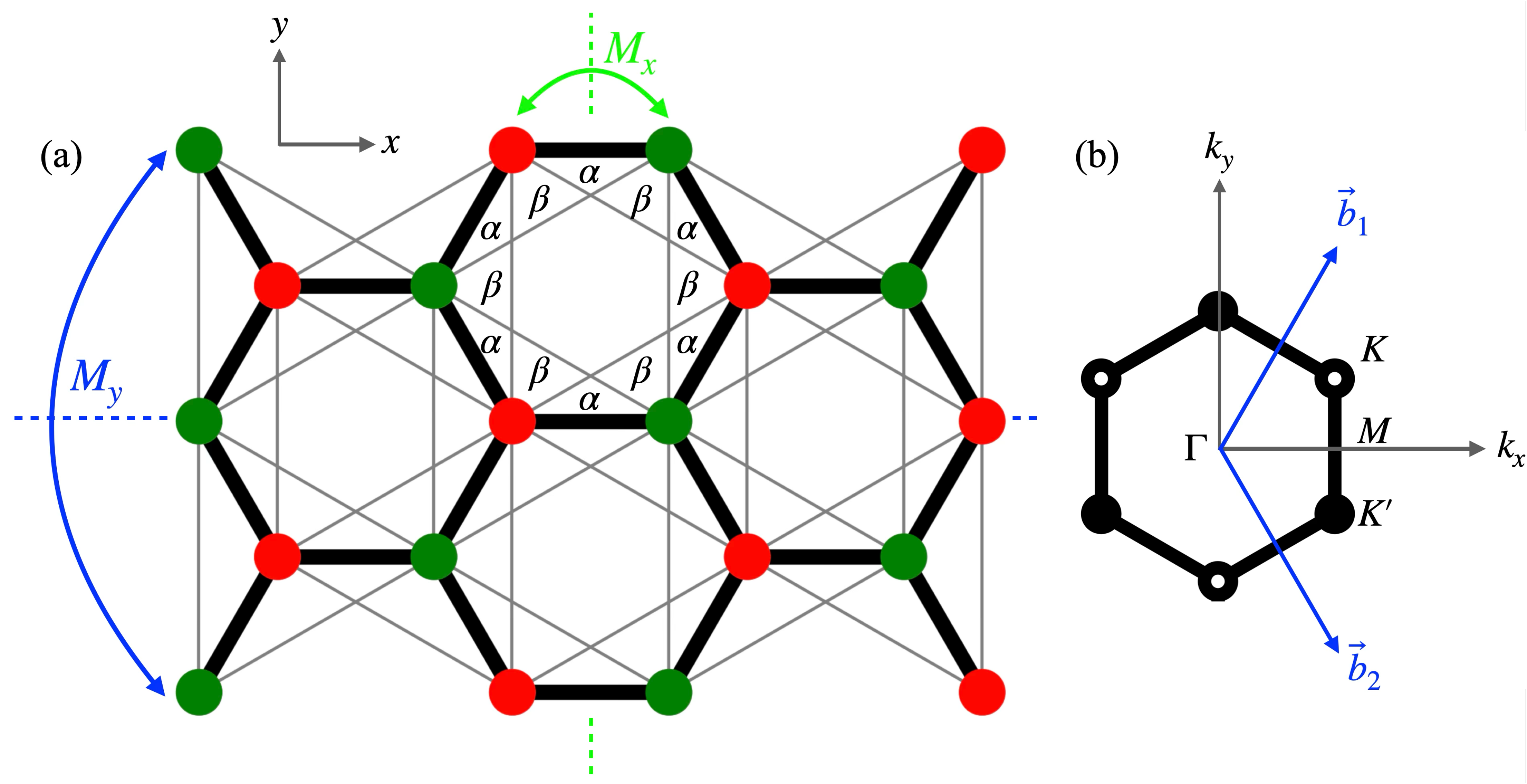}
	\caption{(a) The honeycomb lattice structure. The red and green dots denote two sublattices. The light green (blue) dashed line denotes the mirror plane for $\mathcal{M}_x$ ($\mathcal{M}_y$). The light green double arrow indicates the exchange of sublattices under $\mathcal{M}_x$. The blue double arrow illustrates that the sublattice is mirrored to the same sublattice under $\mathcal{M}_y$. The magnetic flux is denoted by $\alpha, \beta$ and only shown in the middle upper hexagon for brevity. (b) The Brillouin zone for the Haldane model. The reciprocal lattice vectors are $\vec{b}_1=\frac{2\pi}{3}(1,\sqrt{3})$ and $\vec{b}_2=\frac{2\pi}{3}(1,-\sqrt{3})$. %The dashed green box denotes the range used in the numerical integration which is $0<k_x<\frac{4\pi}{3}$ and $0<k_y<\frac{2\pi}{\sqrt{3}}$. 
		The high symmetry point $\bm{K}=(\frac{2\pi}{3}, \frac{2\pi}{3\sqrt{3}})$ is denoted by the empty dot, and $\bm{K'}=(\frac{2\pi}{3}, -\frac{2\pi}{3\sqrt{3}})$ is denoted by the solid black dot. }
	\label{fig:lattice}
\end{figure}

The Haldane model is constructed on a hexagonal lattice of which the unit cell contains two sublattices, as shown in Fig. \ref{fig:lattice}. The lattice vectors are $\vec{a}_1=(\frac{3}{2}\hat{x}+\frac{\sqrt{3}}{2}\hat{y})a_0, \vec{a}_2=(\frac{3}{2}\hat{x}-\frac{\sqrt{3}}{2}\hat{y})a_0$, where $a_0$ is the distance between nearest neighbors. The vectors between the nearest neighbors are $\vec{\delta}_1=-a_0\hat{x}, \vec{\delta}_2=(\frac{1}{2}\hat{x}+\frac{\sqrt{3}}{2}\hat{y})a_0, \vec{\delta}_3=(\frac{1}{2}\hat{x}-\frac{\sqrt{3}}{2}\hat{y})a_0 $.  The vectors between the next nearest neighbors are $\vec{v}_1=-\sqrt{3}a_0\hat{y}, \vec{v}_2=(\frac{3}{2}\hat{x}+\frac{\sqrt{3}}{2}\hat{y})a_0, \vec{v}_3=(-\frac{3}{2}\hat{x}+\frac{\sqrt{3}}{2}\hat{y})a_0$. In the following calculations, we let $a_0=1$ for simplicity. The hopping integral between the (next) nearest neighbors are denoted by $t_1$ ($t_2$). A local magnetic flux is inserted and staggered in a way that the total magnetic flux over a unit cell is zero. The staggered flux is schematically denoted by $\alpha, \beta$ in Fig. \ref{fig:lattice}, where the flux $\alpha+\beta=0$. For brevity, the magnetic flux is labeled in one hexagon in the figure, but present periodically in every hexagon. As a result, the hopping between the next nearest neighbors gains a phase, $\phi=2\alpha+\beta$. In addition, an on-site mass term $M$ that differentiates between sublattices is considered. 

As presented by the green arrows in Fig. \ref{fig:lattice}, the mirror symmetry operation $\mathcal{M}_x:x\rightarrow -x$ interchanges the sublattices. Thus, when the atoms on the sublattices are not equivalent, e.g. by a nonzero mass term, inversion symmetry and $\mathcal{M}_x$ are broken. In contrast, the mirror symmetry $\mathcal{M}_y:y\rightarrow -y$ remains preserved, irrespective of the types of atoms on the sublattices, because the sublattice is always mirrored to the same site. 

The addition of the magnetic flux $\phi$ not only breaks time-reversal symmetry, but also breaks $\mathcal{M}_y$ symmetry because the magnetic flux is an out-of-plane pseudovector and flipped under $\mathcal{M}_{x,y}$ mirror operation. Nonetheless, followed by the time-reversal operation which reverses the magnetic flux again, the system is invariant under $\mathcal{M}_y\mathcal{T}$ operation \cite{Lahiri2024}. 

After Fourier transforming the tight-binding Hamiltonian to momentum space, we obtain 
\begin{eqnarray}\label{eq:ham}
	\begin{split}
		H(k)&=f_0(\vec{k})\sigma_0+f_x(\vec{k})\sigma_x+f_y(\vec{k})\sigma_y+f_z(\vec{k})\sigma_z, \\
		f_0(\vec{k})&=2t_2\cos\phi\left[\cos(\sqrt{3}k_y)+2\cos(\frac{3k_x}{2})\cos(\frac{\sqrt{3}k_y}{2})\right],\\
		f_x(\vec{k})&=t_1\left[\cos k_x+2\cos(\frac{k_x}{2})\cos(\frac{\sqrt{3}k_y}{2})\right],\\
		f_y(\vec{k})&=-t_1\left[\sin k_x-2\sin(\frac{k_x}{2})\cos(\frac{\sqrt{3}k_y}{2})\right],\\
		f_z(\vec{k})&=M+2t_2\sin\phi\left[2\cos(\frac{3k_x}{2})\sin(\frac{\sqrt{3}k_y}{2})-\sin(\sqrt{3}k_y)\right],\\
	\end{split}
\end{eqnarray}
where $\sigma_{x,y,z}$ are Pauli matrices and $\sigma_0$ is identity matrix in the sublattice basis. The energy dispersion is $E(\vec{k})_{\pm}=f_0(\vec{k})\pm\sqrt{\sum_{j=x,y,z}f^2_j(\vec{k})}$ and the energy gap is $2(M\pm 3\sqrt{3}t_2\sin\phi)$, where the positive (negative) sign refers to the gap at $K'(K)$ point. When $\phi=\pm\pi/2$, $f_0$ vanishes, the Hamiltonian is chiral symmetric and energy bands follow $E_+(\vec{k})=-E_-(\vec{k})$. 
The symmetry operations can be represented by Pauli matrices in the basis of sublattices, $\mathcal{M}_x=\sigma_x, \mathcal{M}_y=\sigma_0$, and $T=i\mathcal{K}$, where $\mathcal{K}$ denotes complex conjugate.
It can be shown that the Eq. \ref{eq:ham} is invariant under $\mathcal{M}_y\mathcal{T}$  % $\phi$ and $M$ are both present in the system
\begin{eqnarray}
	(\mathcal{M}_y\mathcal{T})H(k_x,k_y)(\mathcal{M}_y\mathcal{T})^{-1}=H(-k_x,k_y). 
\end{eqnarray}
%Therefore, it is expected that the separation of responses can be observed in the Haldane model. 
Furthermore, the Haldane model possesses $C_{3z}$ symmetry, as the lattice is invariant after $2\pi/3$ rotation. 

The topological phase diagram, characterized by Chern number ($\mathbf{C}$), is shown in the Appendix \ref{sec:app}. 
In the following sections, we discuss the independent components of the second-order conductivities constrained by the lattice symmetry and present the numerical results.              

\begin{figure}
	\includegraphics[width=0.48\textwidth]{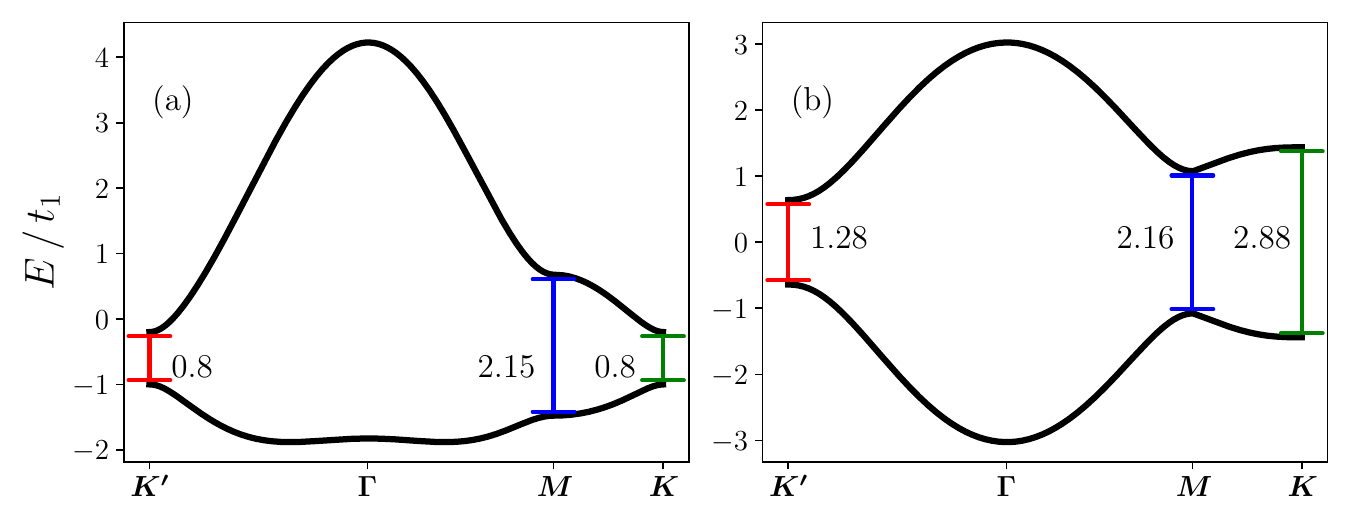}
	\caption{Energy dispersion along $\bm{K'-\Gamma-M-K}$ for $\phi=0$ (a) and $\phi=-\pi/2$ (b). The numerical values indicate the energy gaps at $\bm{K',M,K}$ in the figure. }%(c,d) The $xxx$ component of the shift conductivity for $\phi=0$ and $phi=\pi/2$, respectively. The vertical dashed lines correspond to the energy gaps annotated in (a,b).}
	\label{fig:bands}
\end{figure}
\subsection{The bulk photovoltaic effect}

The d.c. response of the second-order photoconductivies \cite{Sipe2000} are characterized into two processes, injection and shift current. The injection (shift) refers to the change of group velocity (position) during the interband transition.

The shift photoconductivity is given by \cite{Aversa1995,Ahn2020}
\begin{eqnarray}
	\sigma^{c,ab}_{shift}=\frac{-\pi e^3}{\hbar^2}\int\frac{d^dk}{(2\pi)^d} \sum_{n,m}f_{nm}I^{c,ab}_{mn} \delta(\omega_{mn}-\omega)
	%	\sigma_{inj}^{c,ab}=\frac{-2\tau\pi e^3}{\hbar^2}
	%	\int\frac{d^dk}{(2\pi)^d}
	%	 \sum_{nm}f_{nm}I^{c,ab}_{i,mn} \delta(\omega_{mn}-\omega),
	\label{eq:shiftcond}
\end{eqnarray}
where 
$\hbar\omega_{mn}=E_m-E_n$ is the energy difference between two bands, $d$ is the spatial dimension, $f_{nm}=f_n-f_m$, where $f_{n,m}$  is the Fermi-Dirac distribution. The electron charge is $-e$ and $e>0$. 
The integrand for shift conductivity is 
\begin{eqnarray}
	I_{mn}^{c,ab}&=&(R_{mn}^{c,a}-R_{nm}^{c,b})r^b_{nm}r^a_{mn},
	\label{eq:shiftintgrand}
\end{eqnarray}
where $R_{mn}^{c,a}$ is the shift vector \footnote{We adopt the definition from \cite{Ahn2020}, where the shift conductivity was derived accounting for circularly and linearly polarized light, allowing the shift vector to be complex. This definition is slightly different from that in \cite{Sipe2000}, where only the linearly polarized light was considered. } 
\begin{eqnarray}
	R_{mn}^{c,a}&=&r_{mm}^c-r_{nn}^c+i\partial_c {\rm log\ } r_{mn}^a
	\label{eq:shiftvector}
\end{eqnarray}
and $r_{mn}^a=\langle m| i\partial_a|n\rangle$ is the Berry connection. The shift vector is gauge invariant. 
The term  $r^b_{nm}r^a_{mn}$ is the band-resolved quantum geometric tensor, defined as $Q^{ba}=\sum_{n\in\text{occ}}\sum_{m\in\text{unocc}}r_{nm}^br_{mn}^a$ \cite{Provost1980}, where (un)occ denotes the (un)occupied bands. The real part of $Q^{ba}$ is the quantum metric $g^{ba}$, while the imaginary part is proportional to Berry curvature $\Omega^{ba}$. The relation is 
\begin{eqnarray}
	Q^{ba}=g^{ba}-\frac{i}{2}\Omega^{ba}.
	\label{eq:qgt}
\end{eqnarray} 
Eq. \ref{eq:shiftintgrand} can also be written as $i(C_{nm}^{bca}-C_{nm}^{acb})$, where $r_{mn,c}^a=\partial_cr_{mn}^a-i(r_{mm}^c-r_{nn}^c)r_{mn}^a$ and $C_{nm}^{bca}=r_{nm}^br_{mn,c}^a$ is a Hermitian connection \cite{Ahn2022}. 
For numerical calculations, $C_{nm}^{bca}$ is written in terms of the velocity operators and double derivatives of the Hamiltonian
\begin{eqnarray}
	C_{nm}^{bca}&=&\frac{v_{nm}^b}{\omega_{mn}^2}
	\bigg[
	w_{mn}^{ac}-\frac{v_{mn}^c\Delta_{mn}^a+v_{mn}^a\Delta_{mn}^c}{\omega_{mn}}\nonumber\\
	&+&\sum_{p\neq m,n}
	\left(
	\frac{v_{mp}^c v_{pn}^a}{\omega_{mp}}-
	\frac{v_{mp}^a v_{pn}^c}{\omega_{pn}}
	\right)
	\bigg],
	\label{eq:shiftnum}
\end{eqnarray} 
where $w_{mn}^{ac}=\hbar^{-1}\langle m|\frac{\partial^2H}{\partial k_a\partial k_c}|n\rangle$, 
$v_{mn}^a=\hbar^{-1}\langle m|\frac{\partial H}{\partial k_a}|n\rangle$, 
$\Delta_{mn}^a=v_{mm}^a-v_{nn}^a$.  
The last summation in Eq. \ref{eq:shiftnum} is virtual transition which vanishes for two-band models. 
 As shown in \cite{Ahn2020}, the shift vector can be obtained by 
\begin{equation}
	R_{mn}^{c,a}=iC_{nm}^{bca} / (r_{nm}^br_{mn}^a),
\end{equation}
when $r_{nm}^br_{mn}^a$ is nonzero. This expression is equivalent to Eq. \ref{eq:shiftvector}.

The injection conductivity is given by
\begin{eqnarray}
	\sigma^{c,ab}_{inj}=-\tau\frac{2\pi e^3}{\hbar^2}\int\frac{d^dk}{(2\pi)^d} \sum_{nm}f_{nm}D_{mn}^{c,ab}
	\delta(\omega_{mn}-\omega), \nonumber\\
	\label{eq:injcond}
\end{eqnarray}
where $D_{mn}^{c,ab}=\Delta_{mn}^c r_{nm}^br_{mn}^a$ and $\tau$ is the relaxation time. 

The real (imaginary) parts of Eq. \ref{eq:shiftcond} and Eq. \ref{eq:injcond} are the responses to the linearly (circularly) polarized light, dubbed as linear (circular) shift and linear (circular) injection conductivity, respectively \cite{Ahn2020}. 

In the numerical calculation, the Dirac delta function in the equations is replaced with the Lorentzian function 
\begin{eqnarray}
	\mathcal{L}=\frac{1}{\pi}\frac{\gamma/2}{(\omega_{mn}-\omega)^2+(\gamma/2)^2}, 
\end{eqnarray}
where $\gamma$ is the broadening and taken to be $0.04$ in our calculations. Other parameters for the Haldane model used in the calculations: $M=0.4t_1,t_2=0.2t_1$ and $t_1=1$, unless otherwise stated. The energy dispersion for $\phi=0$ and $\phi=0.5\pi$ are shown in Fig. \ref{fig:bands} (a) and (b), respectively.

\subsection{Symmetry analysis}
%In this paper, we study the response driven by linearly polarized light and show the separation of responses in the Haldane model. 
The consequences of the lattice symmetry on the second-order conductivities are given in this section. First, for a system with $\mathcal{M}_k\mathcal{T}$ symmetry, where $\mathcal{M}_k:k\rightarrow-k$, the velocity matrix element has the relationship
%\begin{eqnarray}
$	(\mathcal{M}_kT)v^i_{nm}(k_x,k_y)=(-1)^{\delta_{ik}+1}v^i_{nm}(-k_x,k_y).$
%\end{eqnarray}
As shown by Eq. \ref{eq:injcond}, the symmetry properties of $D^{c,ab}_{mn}$ can be determined by that of $v_{nm}^{b}v^{c}_{mn}v^{a}_{mn}$ and 
\begin{widetext}
	\begin{eqnarray}
		(\mathcal{M}_kT)D^{c,ab}_{mn}(k_x,k_y)=(-1)(-1)^{\delta_{kc}+\delta_{ka}+\delta_{kb}}D^{c,ab*}_{mn}(-k_x,k_y).
	\end{eqnarray} 
\end{widetext}
where the first $(-1)$ on the right-hand-side of the equation is a result of the time-reversal operation on the odd number of velocity operators.
Therefore, the real (imaginary) part of $D^{c,ab}_{mn}$ is an odd function in the first Brillouin zone when $k$ appears even (odd) number of times in the component ${c,ab}$.
Similarly, as shown by Eq. \ref{eq:shiftintgrand} and \ref{eq:shiftnum}, the symmetry properties of $I^{c,ab}_{mn}$ can be determined by that of $iv_{nm}^{b}v^{c}_{mn}v^{a}_{mn}$. The real and imaginary parts of $I^{c,ab}_{mn}$ are interchanged when compared to $D^{c,ab}_{mn}$.  
%\begin{widetext}
%\begin{eqnarray}
%(\mathcal{M}_kT)I^{c,ab}_{mn}(k_x,k_y)(\mathcal{M}_kT)^{-1}=(-1)^{\delta_{kc}+\delta_{ka}+\delta_{kb}}I^{c,ab*}_{mn}(-k_x,k_y),
%\end{eqnarray} 
%\end{widetext}
%where the complex conjugation by the time-reversal operation produces an addition  $(-1)$ on the right-hand-side of the equation. 
Thus, the real (imaginary) part of $I^{c,ab}_{mn}$ is an odd function in the first Brillouin zone when $k$ appears odd (even) number of times in the component ${c,ab}$.
%When $I^{c,ab}_{mn}$ or $D^{c,ab}_{mn}$ is constrained by $\mathcal{M}_k\mathcal{T}$ symmetry to be an odd function in the first Brillouin zone, 
Consequently, the corresponding components of the shift or injection conductivity vanish. The results are summarized in Table \ref{tab:sym}
%\begin{widetext}
	\begin{table*}[t] % [h!] tries to place the table "here" if possible, otherwise at top of page
		\centering % Centers the table content if it's narrower than \linewidth, but with X columns, it should fill
		\caption{The parity of the conductivity tensor after the $\mathcal{M}_k\mathcal{T}$ operation. The sign $+$ and $-$ denote the even and odd parity, respectively. The response vanishes for odd parity. } % Essential for tables
		\label{tab:sym}
		\begin{tabularx}{\linewidth}{L CCCC} % L for the first column (left-aligned, stretching), C for the other four (centered, stretching)
			\toprule % From booktabs for a professional top line
			Response & linear injection & circular injection & linear shift & circular shift \\
			\midrule % From booktabs for a professional line between header and content
			even number of $k$ in $cab$ & $-$ & $+$ & $+$ & $-$ \\
			odd number of $k$ in $cab$ & $+$ & $-$ & $-$ & $+$ \\
			\bottomrule % From booktabs for a professional bottom line
		\end{tabularx}
	\end{table*}

Next, we turn to the Haldane model as an example. 
%Interestingly, in the presence of $\mathcal{MT}$ or $\mathcal{C}_2\mathcal{T}$ symmetry, certain components of the conductivity tensors show $T$-symmetric responses, while others show $\mathcal{PT}$- symmetric responses. This phenomenon is dubbed the separation of responses in Ref \cite{Ahn2020}. 
As a results of $\mathcal{M}_y\mathcal{T}$ symmetry of the Haldane model, the component with even number of $y$ is time-reversal symmetric and linear shift conductivity is allowed. In contrast, the component with odd number of $y$ is $\mathcal{MT}$ symmetric and linear injection is allowed. 

In addition, the Haldane model possesses $C_{3z}$ symmetry. By Neumann principle, there are only two independent components of the second-order response tensor \cite{Boyd,Murti2021Physics}. Namely, 
\begin{eqnarray}
	-\sigma^{x,xx}&=&\sigma^{y,xy}=\sigma^{y,yx}=\sigma^{x,yy}\nonumber\\
	-\sigma^{y,yy}&=&\sigma^{x,yx}=\sigma^{x,xy}=\sigma^{y,xx}.
	\label{eq:neumann}
\end{eqnarray}

Constrained by $C_{3z}$ that the conductivity is symmetric under interchaning $b$ and $c$, i.e. $\sigma^{y,xy}=\sigma^{y,yx}$ and $\sigma^{x,yx}=\sigma^{x,xy}$, the circular responses vanish in the Haldane model.  
We have numerically examined that all the components of the conductivities obey the symmetry analysis presented in this section. In the following, the independent components, $\sigma^{x,xx}_{shift}$ and $\sigma^{y,yy}_{inj}$, are shown. 
\section{numerical results}\label{sec:results}

\subsection{BPVE and quantum geometry}
\begin{figure}
	\includegraphics[width=0.47\textwidth]{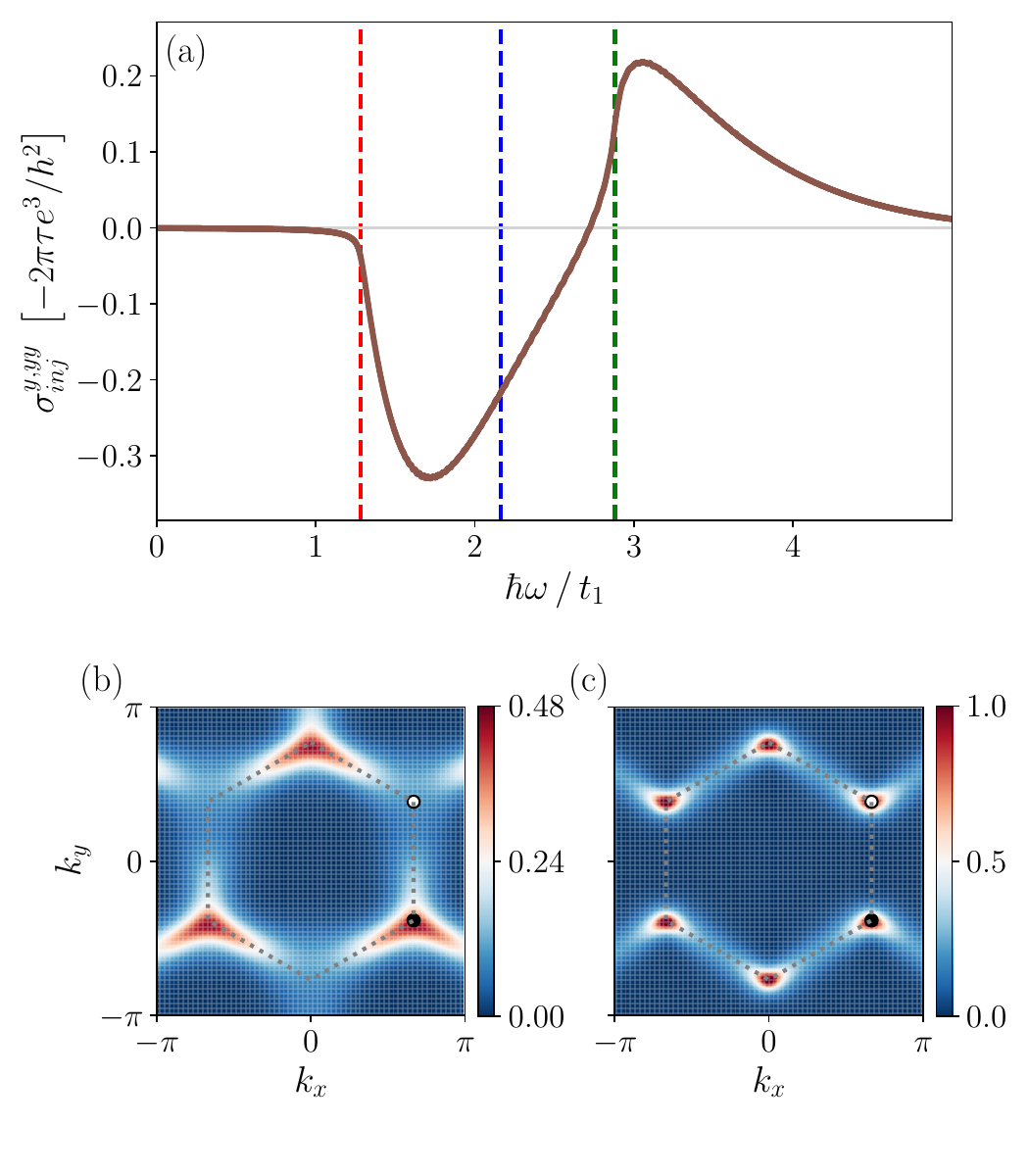}
	\caption{(a) The injection conductivity as a function of photon energy for the Haldane model with $M/t_1=0.4$, $\phi=-\pi/2$ ($\mathbf{C}=-1$) and $\mu=0$. The vertical dashed lines, ordered from left to right, indicate the photon energies corresponding to the $\bm{K' 
		,M,K}$ energy gaps, as annotated in  Fig. \ref{fig:bands}. (b) The momentum resolved quantum metric $g^{yy}$ for the conductivity in (a). (c) The momentum resolved quantum metric $g^{yy}$ for $M/t_1=0.4$, $\phi=0$ ($\mathbf{C}=0$) and $\mu=0$. }
	\label{fig:injmetric}
\end{figure}

%\comm{compute quantum volume...nothing new}
Fig. \ref{fig:injmetric} (a) shows the $yyy$ component of the linear injection conductivity  as a function of photon energy for the Haldane model with $M/t_1=0.4$ and $\phi=-\pi/2$. The Chern number for this set of parameters is one ($\mathbf{C}=1$). The chemical potential $\mu$ is chosen to be $0$, inside the gap. Since the linear injection conductivity vanishes under time-reversal symmetry, only the result for $\phi\neq 0$ is shown. The vertical dashed lines indicate the photon energies equivalent to the energy gaps at high symmetry points, annotated in Fig. \ref{fig:bands} (b). The onset frequency corresponds to the energy gap at $\bm{K'}$. The response is nonvanishing at higher photon frequencies corresponding to larger energy gaps at $\bm{M}$ and $\bm{K}$ points, as denoted by the blue and green dashed lines, respectively. Fig. \ref{fig:injmetric} (b) shows the quantum metric $g^{yy}$, which dominates near $\bm{K'}$ and approaches zero near $\bm{\Gamma}$. For comparison, the quantum metric ($g^{yy}$) for $\phi=0$ ($\mathbf{C}=0$) in the topological trivial phase is shown in Fig. \ref{fig:injmetric} (c). In contrast to Fig. \ref{fig:injmetric} (b), $g^{yy}=|r_{12}^{y}|^2$ is $\mathcal{M}_y$ symmetric, reflecting $\mathcal{M}_y$ symmetry of the Haldane Hamiltonian when $\phi=0$, as mentioned in sec. \ref{sec:hamiltonian}.

\begin{figure}
	\includegraphics[width=0.5\textwidth]{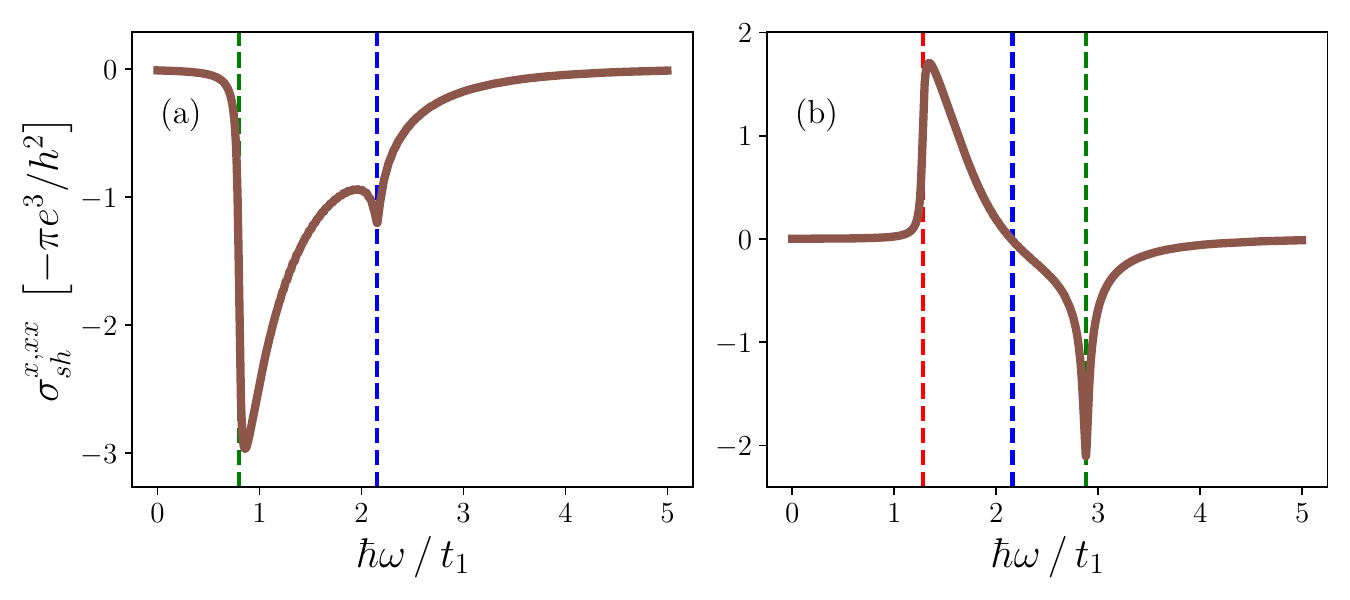}
	\caption{The $xxx$ component of the shift conductivity for (a) $\phi=0$ and $\mu=-0.5$ and (b) $\phi=-\pi/2$ and $\mu=0$. $M/t_1=0.4$ for both panels. The color-coded vertical dashed lines correspond to the energy gaps annotated in Fig. \ref{fig:bands}. }
	\label{fig:sh}
\end{figure}

Fig. \ref{fig:sh} shows the $xxx$ component of the linear shift conductivity tensor as a function of photon energy. Fig. \ref{fig:sh} (a) shows the result for $\phi=0$ and $\mu=-0.5$, due to time-reversal symmetry, $\bm{K}$ and $\bm{K'}$ are degenerate, leading to a strong peak near $\hbar\omega=0.8$, as indicated by the green dashed line. A second peak is present at higher energy, showing the resonance at the energy equivalent to the gap at $\bm{M}$, as indicated by the blue dashed line. Fig. \ref{fig:sh} (b) shows the result for $\phi=-\pi/2$ and $\mu=0$, where time-reversal symmetry is broken. %The response at $\bm{K'}$ flips sign, as a consequence of significant sign change of shift vectors led by band inversion near $\bm{K'}$ \cite{Qian2022}. 
When the photon energy approaches the energy gap at $\bm{M}$ point, the shift conductivity is negligible. 

\begin{figure}
	\includegraphics[width=0.5\textwidth]{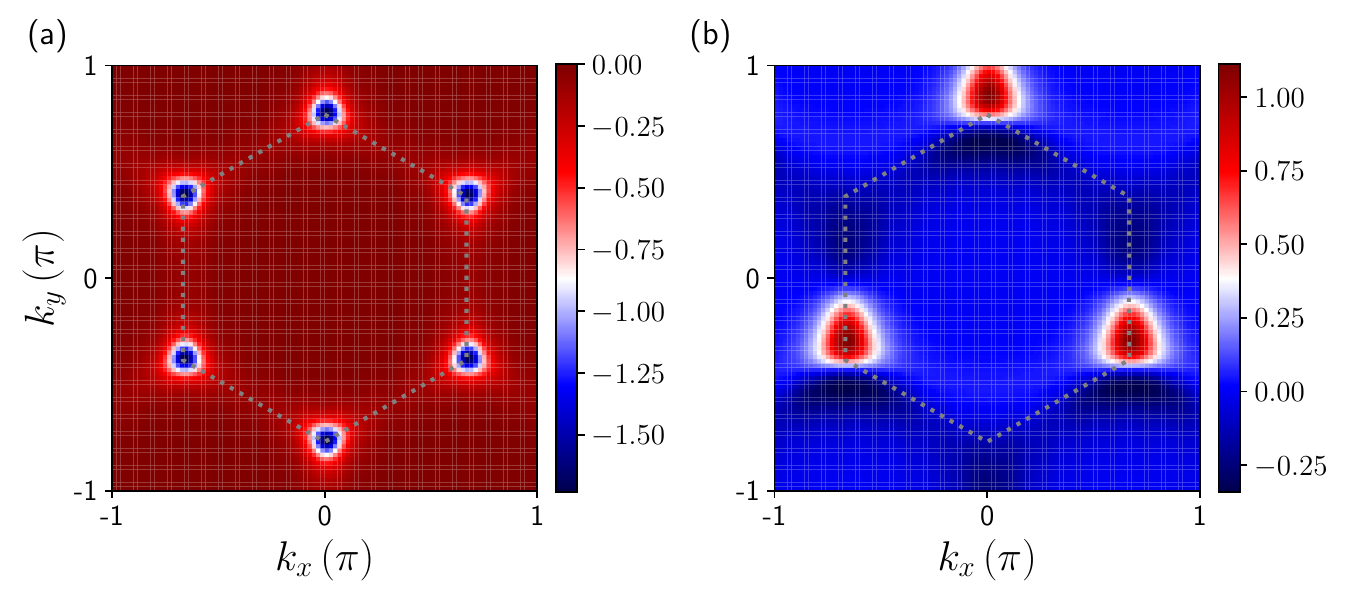}
	\caption{(a) The symplectic connection $-\text{Im}[C_{12}^{x,xx}]$ for $\phi=0$ and $M/t_1=0.4$. (b) The symplecitc connection $-\text{Im}[C_{12}^{x,xx}]$ for $\phi=-\pi/2$ and $M/t_1=0.4$. }
	\label{fig:shc}
\end{figure}

The shift conductivity is related to the Hermitian connections, whose real and imaginary parts are the metric and symplectic connections, respectively \cite{Ahn2022,Hsu2023}. For the $xxx$ component, only the imaginary part of the connection (symplectic connection) contributes to the shift conductivity since $I_{mn}^{c,ab}=i(C_{nm}^{bca}-C_{mn}^{acb})$. 
Fig. \ref{fig:shc} shows the momentum resolved symplectic connections for the Haldane models. When $\phi=0$, the system is in the topological trivial phase [Fig. \ref{fig:shc} (a)], the symplectic connections are the same at $\bm{K}$ and $\bm{K'}$, in sharp contrast to Berry curvature shown in Fig. \ref{fig:berry} (a). When $\phi=-\pi/2$ and in the topological phase, [Fig. \ref{fig:shc} (b)], the symplectic connection dominates near $\bm{K'}$. The signs near $\bm{K'}$ and $\bm{K}$ are opposite to each other, in agreement with the opposite resonant responses when the photon energies are equivalent to the energy gaps at $\bm{K'}$ and $\bm{K}$, as shown in Fig. \ref{fig:sh} (b). 
The sign difference near $\bm{K'}$ between Fig. \ref{fig:shc} (a) and (b)  can be understood by band inversion led by topological phase transition. In Eq. \ref{eq:avgsh}, the term $g_{12}^{xx} = |r_{12}^x|^2$ is always non-negative. Consequently, any observed sign change in ${C^{x,xx}_{12}}$ must originate from the shift vectors. This sign change in the shift vector is the expected result of band inversion during the system's transition from the trivial to the topological phase. %\cite{Qian2022}.%, i.e. $R_{vc}^{c,a}$ is flipped to $R_{cv}^{c,a}=-R_{vc}^{c,a}$ in the interted regime, where $v(c)$ denotes the eigenstate of the valence (conduction) band in the normal regime. %\cite{Qian2022}. %The sign change of the shift vector across topological phase transition has also been shown in the Kane-Mele model \cite{Qian2022}.}

\subsection{Zone-averaged Responses and symmetry breaking} %/ The Role of Symmetry Breaking: Analysis of Zone-Averaged Responses}
To explore the effect of the strength of inversion symmetry breaking by $M$ and the time reversal symmetry breaking by $\phi$ when the lower energy band is fully occupied, we define the zone-averaged responses by integrating $I^{x,xx}_{12}$ and $D^{y,yy}_{12}$ in the first Brillouin zone, specifically, 
\begin{eqnarray}
	\overline{I^{x,xx}_{12}}&=&\int d^2kI_{12}^{x,xx}=\int d^2k(R^{x,x}_{12}-R^{x,x}_{21})g_{12}^{xx}\label{eq:avgsh}\\
	\overline{D^{y,yy}_{12}}&=&\int d^2kD_{12}^{y,yy}=\int d^2k\Delta^{y}_{12}g_{12}^{yy}.\label{eq:avginj} 
\end{eqnarray}
This is equivalent to the integrated response over all frequencies, up to a factor of $-\frac{\pi e^3}{h^2}$ for shift conductivity, and $-\tau\frac{2\pi e^3}{h^2}$ for injection conductivity. The integral over frequency is reasonable because applications often involve light incident over a range of frequencies. Eq. \ref{eq:avgsh} and \ref{eq:avginj} model the incidence by light with flat broad spectrum \cite{Rangel2017,Fregoso2017}. %The exclusion of $\delta(\omega_{mn}-\omega)$ ensures that every $k$ point is taken into account. 
%The results of the integration are denoted by $\overline{I^{x,xx}_{12}}$ and $\overline{D^{y,yy}_{12}}$, respectively. 

\begin{figure}
	\includegraphics[width=0.5\textwidth]{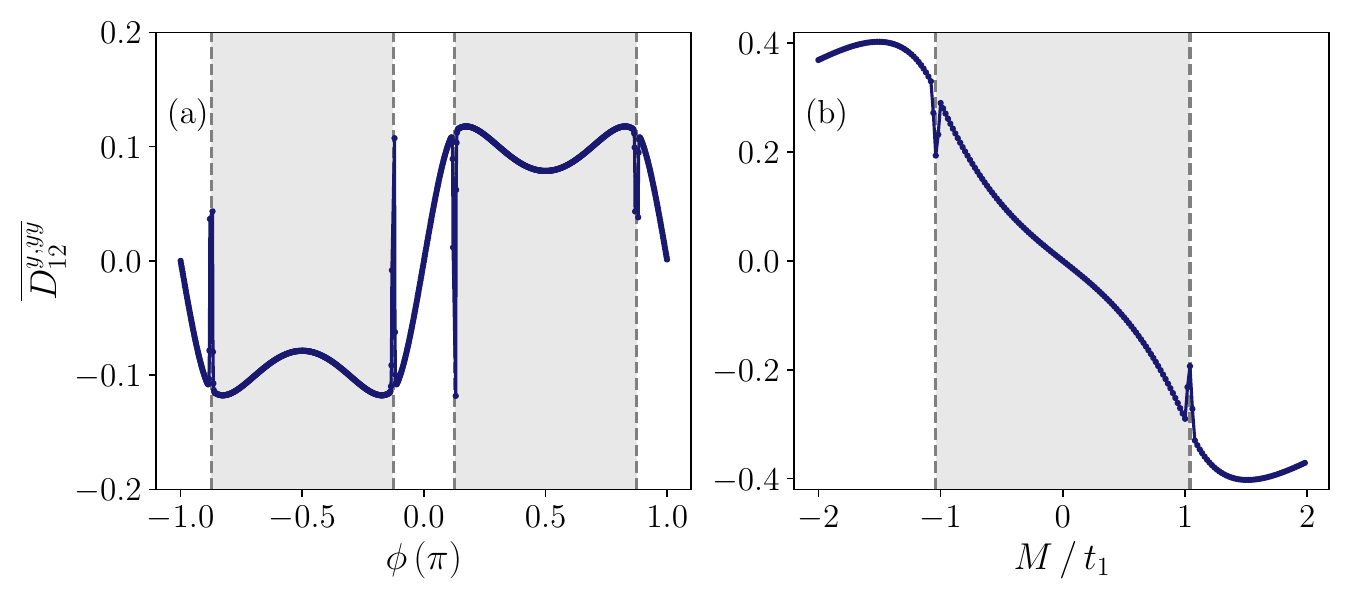}
	\caption{(a) $\overline{D_{12}^{y,yy}}$ as a function of $\phi$ for $M=0.4/t_1$. (b) $D_{12}^{y,yy}$ as a function of $M/t_1$ for $\phi=-\pi/2$. The light gray areas are the regimes for topological phases. The dashed vertical lines denote where the energy gap closes. }
	\label{fig:injdep}
\end{figure}

Fig. \ref{fig:injdep} (a) shows $\overline{D^{y,yy}_{12}}$ as a function of $\phi$ with $M/t_1=0.4$. The light gray areas are the regimes of topological phases. The phase boundaries correspond to energy gap closures when $\phi=\arcsin(\pm\frac{M}{3\sqrt{3}t_2})$. For the model parameters used in the calculation, the gap closes when $\phi=\pm 0.126\pi, \pm 0.874\pi$, leading to diverging values on the phase boundaries. Across topological phase transitions, $\overline{D^{y,yy}_{12}}$ remain the same sign, as group velocity does not change sign when band inverts. 
%When $\phi$ is nonzero, $\mathcal{M}_y$ and $\mathcal{T}$ are broken, whereas $\mathcal{M}_y\mathcal{T}$ is preserved, allowing nonvanishing $\sigma_{\text{inj}}^{yyy}$. 
%The results show that $\overline{D^{y,yy}_{12}}$ is nonvanishing when $\phi\neq 0$ and remain the same sign across topological phase transitions. 
%The flux $\phi$ enters $f_0$ and $f_z$ of the Hamiltonian. However, $f_0$ does not contribute to $\overline{D^{y,yy}_{12}}$ because $\langle n|\partial_yf_0I|m\rangle=\delta_{n,m}\partial_yf_0$. %and $\partial_yf_0$ in the velocity expectation values cancel out between bands $\partial_y(E_k=f_0\pm\sqrt{f_i^2})$.  Thus, the oddness of $\overline{D^{y,yy}_{12}}$ in $\phi$ can be understood as a result of $\sin\phi$ in $f_z$.
Fig. \ref{fig:injdep} (b) shows $\overline{D^{y,yy}_{12}}$ as a function of $M$ with $\phi=-\pi/2$. The topological phase boundaries correspond to band gap closures at $M=\pm 1.04$, leading to diverging values. Overall, $\overline{D^{y,yy}_{12}}$ is odd in both $M$ and $\phi$. This is consistent with the symmetry principles for linear injection currents. The response must vanish when $\phi=0$, which restores time-reversal symmetry, or $M=0$, which restores inversion symmetry.

\begin{figure}
	\includegraphics[width=0.5\textwidth]{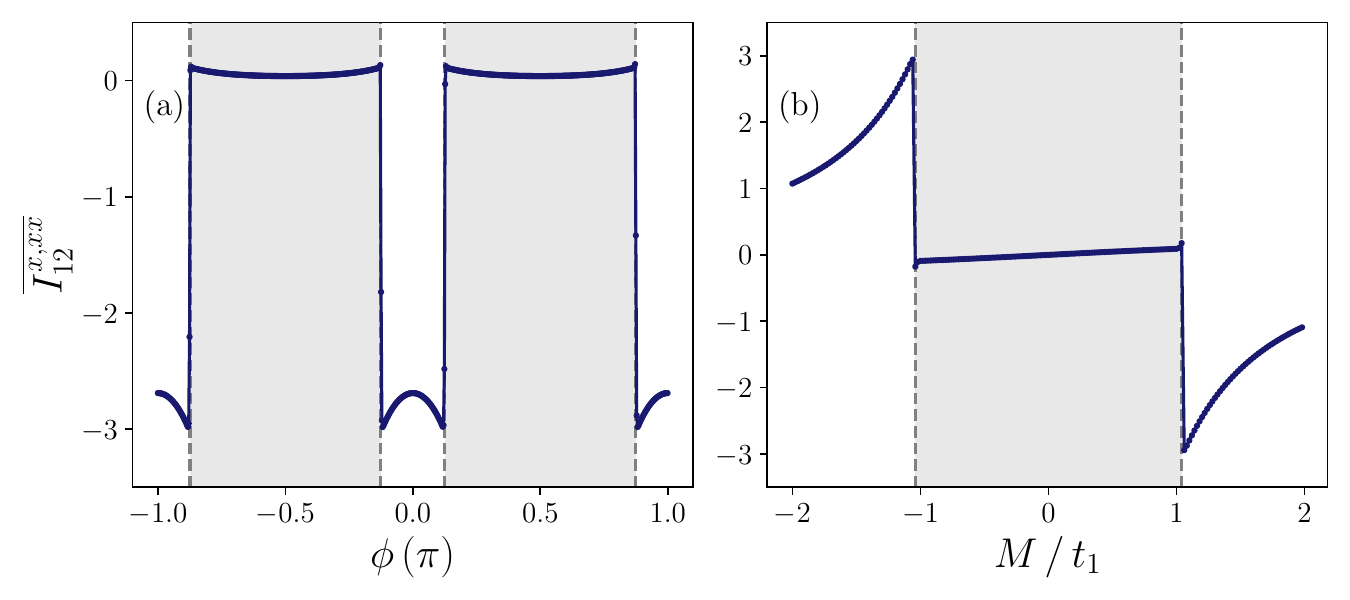}
	\caption{(a) $\overline{I_{12}^{x,xx}}$ as a function of $\phi$. (b) $\overline{I_{12}^{x,xx}}$ as a function of $M/t_1$ for $\phi=-\pi/2$. The light gray areas are the regimes for topological phases. The dashed vertical lines denote where the energy gap closes. }
	\label{fig:shdep}
\end{figure}

Fig. \ref{fig:shdep} (a) and (b) show that $\overline{I^{x,xx}_{12}}$ is an even function of $\phi$ and an odd function of $M$, respectively. This agrees with the symmetry constraint for linear shift currents, as the response vanishes when $M=0$, which restores inversion symmetry. In contrast, shift currents are allowed by time-reversal symmetry, thus explaining the evenness in $\phi$. Furthermore, the figures show that the response changes significantly across the topological phase transitions, a behavior that can be attributed to band inversion. % and the Berry phase effect. 
%Since  $\overline{I^{x,xx}_{12}}=\int d^2k(R^{x,x}_{12}-R^{x,x}_{21})g_{12}^{xx}$ (Eq. \ref{eq:shiftintgrand}) 
%Since in Eq. \ref{eq:avgsh}, $g_{12}^{xx}=r_{12}^{x}r_{21}^x=|r_{12}^x|^2>=0$, the sign change in $\overline{I^{x,xx}_{12}}$ reflects the significant sign change in the shift vectors due to band inversion near $\bm{K'}$ in the topological phase. 
After integration over the Brillouin zone, $\overline{I^{x,xx}_{12}}$ in the topological phase is significantly smaller than that in the trivial phase because shift vectors in the band inversion and normal regimes have opposite signs, leading to cancellation. In contrast, in the trivial phase, each $k$ point contributes constructively, resulting in a substantial $\overline{I^{x,xx}_{12}}$ value. This behavior is consistent across other parameters we calculated (Fig. \ref{fig:otherparam}). We expect this magnitude difference in the zone-averaged shift response between the trivial and topological phases to be a general feature in other models because band-inversion occurs at the topological phase transition.

The discontinuity at the phase boundary is related to the change of Chern number. 
The difference of the shift vectors can be written in terms of Berry connections  $R^{x,x}_{12}-R^{x,x}_{21}=2(r_{11}^x-r_{22}^x+\partial_x\phi_{12}^x)$, where $\phi_{12}^x$ is the phase of the Berry connection $r_{12}^x=|r_{12}^x|e^{-i\phi_{12}^x}$. 
%When gap closes,$r_{12}^x$ vanishes and $\phi_{12}^x$ is ill-defined at certain $k$-points in the Brillouin zone, obstructing the existence of a global smooth gauge \cite{Yoshida2025Diverging}. 
In the trivial phase with $\mathbf{C}=0$, one can find a global gauge such that $\phi_{12}^x$ is zero in the whole Brillouin zone \cite{Fregoso2017, Yoshida2025Diverging}. On the contrary, in the topological phase $(\mathbf{C}=\pm1)$, Berry connections are ill-defined at certain $k$-points in the Brillouin zone, obstructing the existence of a global smooth gauge, leading to a discontinuity at phase transition.
 %The integral $\int \partial_x\phi_{12}^x dk_x dk_y$ is nonvanishing in the topological phase.   %Thus, giving rise to a discontinuity at phase transition.

\begin{figure*}
	\includegraphics[width=\textwidth]{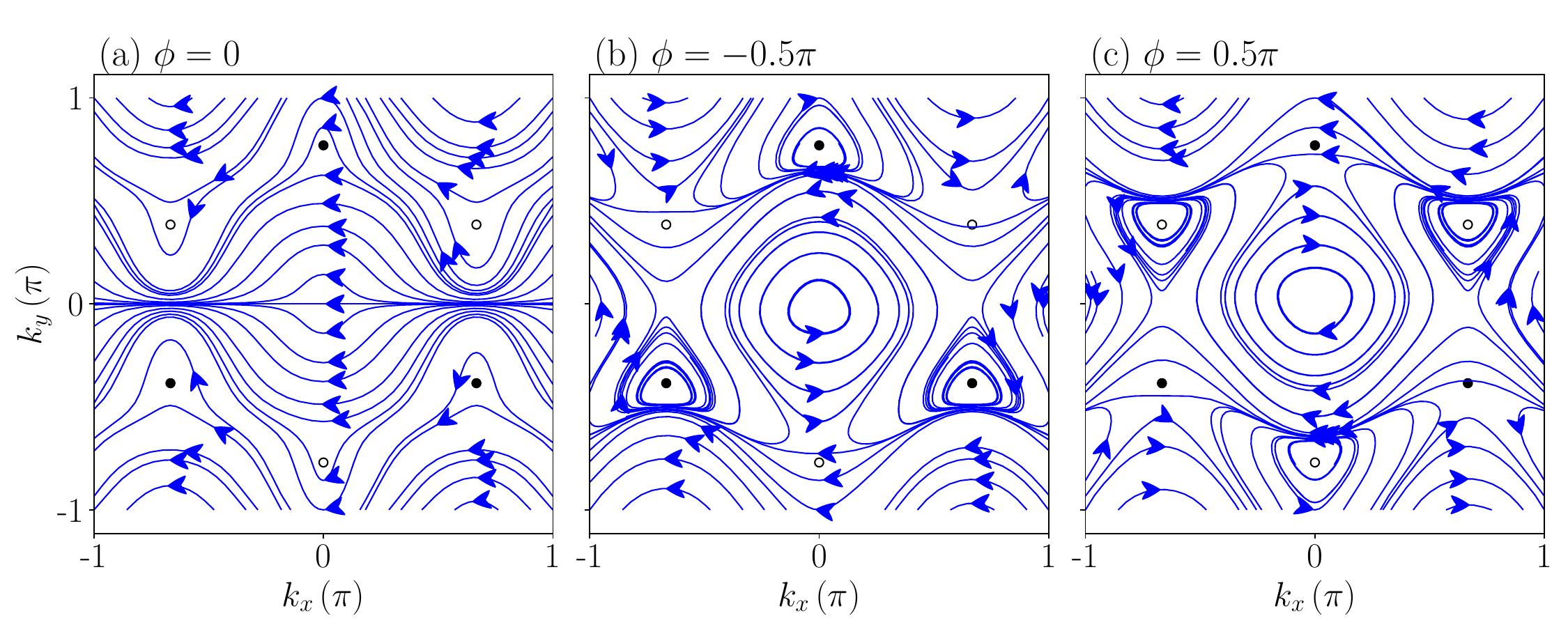}
\caption{
	Stream plot of vector fields of the symplectic connection $(\text{Im}[C_{12}^{x,xx}],\text{Im}[C_{12}^{y,xx}])$% \(\text{Im}(R^{x,x}_{12}g^{xx}_{12}, R^{y,x}_{12}g^{xx}_{12})\) 
	in \(k\)-space for \(M = 0.4/t_1\). %, near the \(\bm{K}\) (top row) and \(\bm{K'}\) (bottom row) points. (a,d) Trivial phase, corresponding to Chern number $\mathbf{C}=0$, with model parameter \(\phi = 0\); (b,e) topological phase with \(\phi = -0.5\pi\), corresponding to \(\mathbf{C} = -1\); (c,f) topological phase with \(\phi = 0.5\pi\), corresponding to \(\mathbf{C} = 1\). White circles and black dots denote the \(\bm{K}\) and \(\bm{K'}\) points, respectively. 
	(a) Trivial phase, corresponding to Chern number $\mathbf{C}=0$, with model parameter \(\phi = 0\); (b) topological phase with \(\phi = -0.5\pi\), corresponding to \(\mathbf{C} = -1\); (c) topological phase with \(\phi = 0.5\pi\), corresponding to \(\mathbf{C} = 1\). White circles and black dots denote the \(\bm{K}\) and \(\bm{K'}\) points, respectively.  
}

	\label{fig:shvec}
\end{figure*}

By visualizing the vector field of symplectic connection  $(\text{Im}[C_{12}^{x,xx}],\text{Im}[C_{12}^{y,xx}])$ in Fig. \ref{fig:shvec}, we found sharp contrast between topological trivial and nontrivial phases. For the trivial phase with $\phi=0$ ($\mathbf{C}=0$) in panel (a), the vector field is $\mathcal{M}_y$ symmetric and vortex-free. In contrast, for $\phi=-0.5\pi$ ($\mathbf{C}=-1$) in panel (b), an counterclockwise vortex is observed near $\bm{\Gamma}$ and a clockwise vortex near $\bm{K'}$; for $\phi=0.5\pi$ ($\mathbf{C}=1$) in panel (c), a clockwise vortex is observed near $\bm{\Gamma}$ and an counterclockwise vortex is observed near $\bm{K}$.

%The discontinuity at the phase boundary is a result of the sign change of $r_{12}^x$, which is proportional to $f_z$ near gap closure as shown in Appendix \ref{app:approx}, leading to an abrupt change of in $\phi_{12}^x$. 
%The jump can be viewed as the reversal of the polarization difference during the optical transition \cite{Fregoso2017}. 
%Under the approximation that $g_{12}^{xx}$ is a constant, $\overline{I^{x,xx}_{12}}\propto\int d^2k(R^{x,x}_{12}-R^{x,x}_{21})$. The integral gives rise to the polarization difference between two bands up to a phase. 
%This jump is reminicent of the shift of the Wannier center in the one-dimensional SSH chain. However, here we consider the shift of the wave packet between bands. 
%Furthermore, in sharp contrast to the response resulting from Berry curvature, the magnitude of $\overline{I^{x,xx}_{12}}$ is larger in trivial regimes. 
%$\overline{I^{x,xx}_{12}}$ is even in $\phi$, as shown in Fig. \ref{fig:shdep} (a).  %The linear shift response is not constrained by time-reversal symmetry. 

%\begin{widetext}
%\end{widetext}

\section{Conclusion}\label{sec:disc}
We study the second-order photoconductivities in two-dimensional mirror-time ($\mathcal{M}\mathcal{T}$) symmetric systems and calculate the quantum geometrical quantities, including quantum metric and symplectic connection for the Haldane model. For a two-dimensional system with $\mathcal{M}_k\mathcal{T}$ symmetry, we show that when the components $cab$ of the second-order photoconductivity tensors contain odd (even) number of $k$, the linear shift (injection) vanishes. 

For the Haldane model, in addition to $\mathcal{M}_y\mathcal{T}$ symmetry, $C_{3z}$ symmetry is preserved, leading to vanishing  circular shift and injection currents. The linear injection current is nonvanishing when time-reversal symmetry is broken. It is intrinsically related to the quantum metric, exhibiting a significant contribution from one valley and a moderate contribution from the other valley and the $\bm{M}$ point. 
The linear shift conductivity shows strong resonance when the photon energy is close to the energy gap at $\bm{K}$ or $\bm{K'}$ valley. It is intrinsically related to symplectic connection, which possesses the same signs for both valleys when $\phi=0$, in sharp contrast to Berry curvature. Across the topological phase transitions, the linear injection conductivity does not change sign, whereas the linear shift conductivity undergoes a sign change. 

The Haldane model is the minimal model of a Chern insulator and can be extended to study the low energy physics of several materials. In this work, we employ the Haldane model to theoretically investigate the bulk photovoltaic effects. Our analysis explores how the model's inherent quantum geometry and symmetries influence the resulting photocurrent responses. The theoretical results can be verified in ultracold atomic systems, which feature tunable symmetry breaking $\cite{Jotzu2014}$ and allow for the simulation of optical responses $\cite{Tran2017}$. This study provides insight for probing the symmetry and quantum geometry in real materials with bulk photovoltaic effects. 
\begin{acknowledgments}
	H.-C. H. would like to thank Dr. Xiao Zhang for insightful discussions. 
	The authors acknowledge the support from the National Science and Technology Council (NSTC) under the grant No. 113-2628-M-004-001-MY3 and the National Center for Theoretical Sciences (NCTS) in Taiwan. 
\end{acknowledgments}

\appendix
\section{Berry curvature and Chern number}\label{sec:app}
The Chern number was computed by
\begin{eqnarray}
	\begin{split}
	\mathbf{C}&=\frac{1}{2\pi}\int \Omega^{xy}(k_x,k_y) d^2k\\
	%\Omega_{xy}(k_x,k_y)&=
	\end{split}
\end{eqnarray}
where $\Omega^{xy}$ is the Berry curvature and equivalent to $-2\text{Im}[Q^{xy}]$ quantum geometric tensor. In the numerical calculation, we have used the identity $r_{nm}^a={\langle n|\frac{\partial H}{\partial_{k_a}}|m\rangle}/{i(E_n-E_m)}$ and 
\begin{eqnarray}
	Q^{xy}=\sum_{n,E_n<E_f}\sum_{m,E_m>F_f}\frac{\langle n|\frac{\partial H}{\partial k_x}|m\rangle \langle m|\frac{\partial H}{\partial k_y}|n\rangle}{(E_m-E_n)^2}.
\end{eqnarray}

\begin{figure}
	\includegraphics[width=0.5\textwidth]{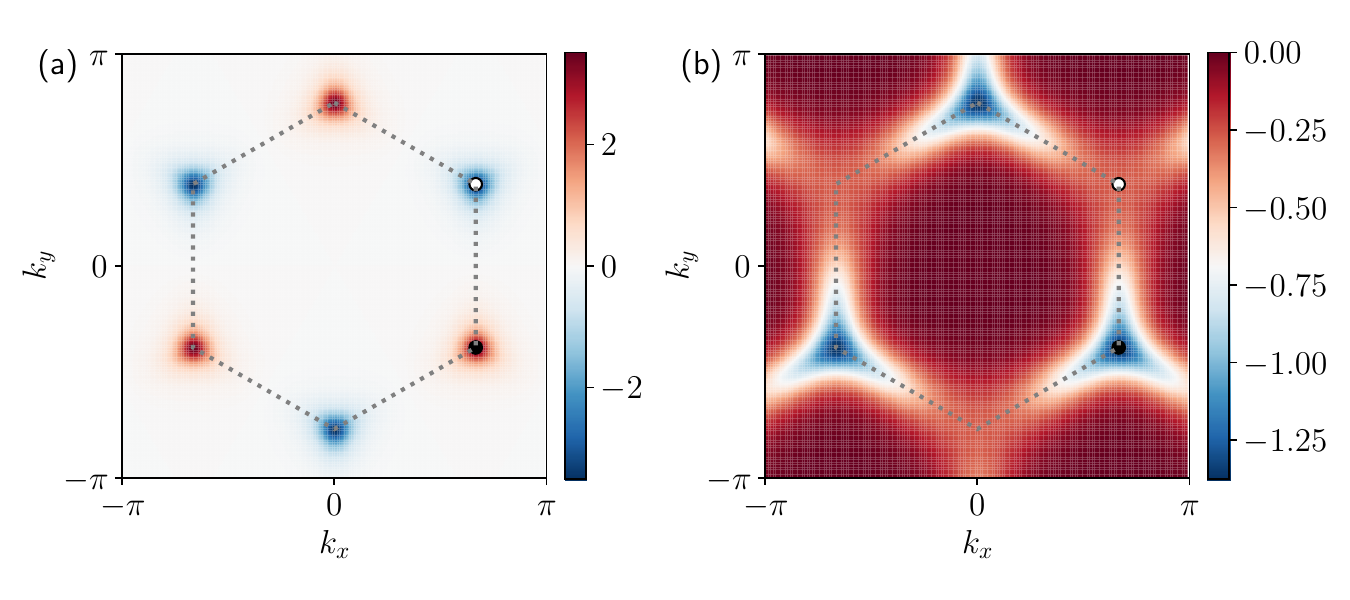}
	\caption{Berry curvature for (a) $\phi=0$  and (b) $\phi=-\pi/2$. Other parameters are $M/t_1=0.4, t_1=1, t_2=0.2$.  }
	\label{fig:berry}
\end{figure}

The Berry curvature distributions in the momentum space for $\phi=0$ and $\phi=-\pi/2$ are shown in Fig. \ref{fig:berry}. When $\phi=0$, the system preserves time-reversal symmetry. The Berry curvatures near $\bm{K}$ and $\bm{K'}$ are opposite. In contrast, when $\phi=-\pi/2$, time-reversal symmetry is broken. The Berry curvature is dominant at $\bm{K}$. 

\begin{figure}
	\includegraphics[width=0.45\textwidth]{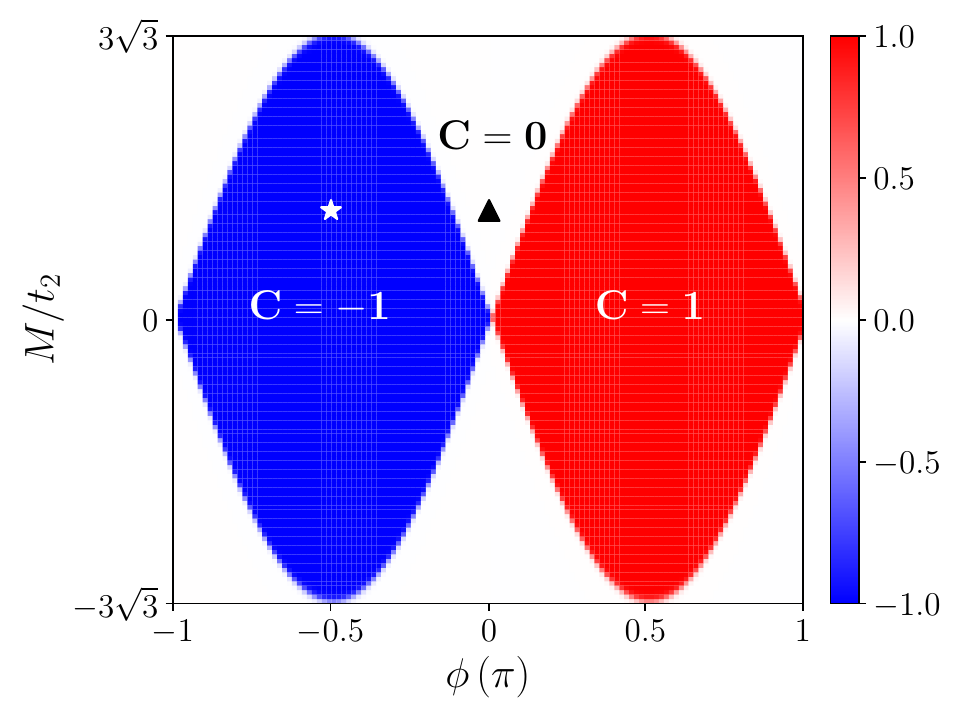}
	\caption{Chern number as a function of $M$ and $\phi$ for the Haldane model. The black triangle ($\blacktriangle$) and white star ($\star$) denote the representative parameters for the topologically trivial ($M=2t_2=0.4$, $\phi=0$) and topological ($M=2t_2=0.4t_1$, $\phi=-0.5\pi$) regimes, respectively. For Figs. \ref{fig:bands}, \ref{fig:sh}, \ref{fig:shc}, and \ref{fig:berry}, the $\blacktriangle$ parameters are used for subfigures (a), and the $\star$ parameters are used for subfigures (b). In Fig. \ref{fig:injmetric}, however, the $\star$ parameters are used for subfigures (a) and (b), while the $\blacktriangle$ parameters are used for subfigure (c). In Fig. \ref{fig:shvec}, the $\blacktriangle$ parameters are used for subfigures (a) and the $\star$ parameters are used for subfigures (b).
		}
	\label{fig:ch}
\end{figure}

The Chern number ($\mathbf{C}$) as a function of $M$ and $\phi$ is shown in Fig. \ref{fig:ch}. 
The phase boundaries correspond to gap closure, determined by the condition $M\pm 3\sqrt{3}\sin\phi$. %In this paper, the definition of positive $\phi$ is opposite to that of the original paper. Thus, the Chern number is also negative as compared to the original paper. 
\section{More data of zone-averaged shift response}
\begin{figure}
	\includegraphics[width=0.5\textwidth]{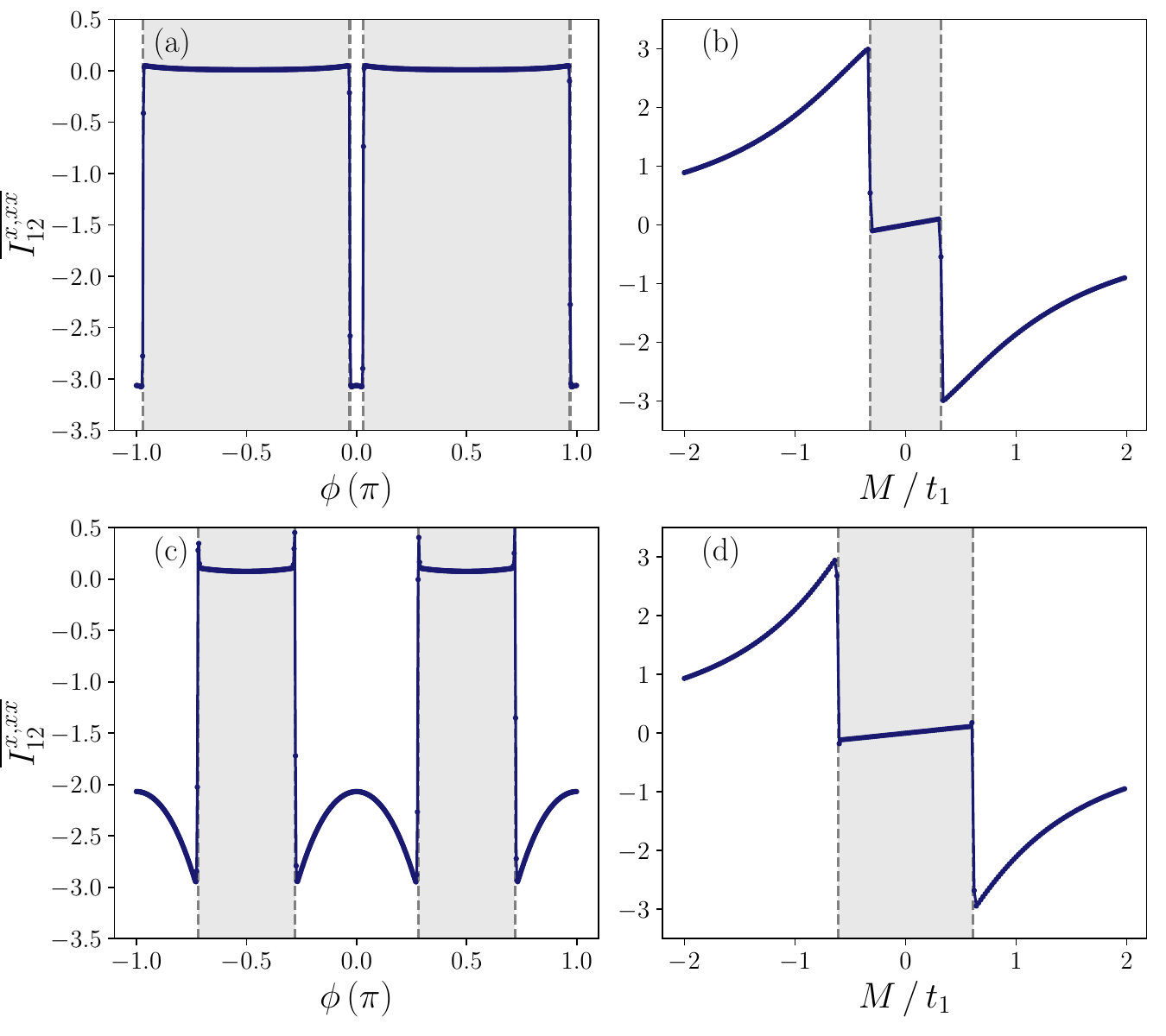}
	\caption{Zone-averaged shift response for (a) $M/t_1=0.1$, (b) $\phi=-0.1\pi$, (c) $M/t_1=0.8$ and (d) $\phi=-0.8\pi$. Other parameters are $t_1=1, t_2=0.2$.  }
	\label{fig:otherparam}
\end{figure}

Fig. \ref{fig:otherparam} presents the zone-averaged shift response (Eq. \ref{eq:avgsh}) for other choices of parameters. The discontinuities at the topological phase boundaries can be observed in all four panels. The origin of the discontinuity and the weak response in the topological phase are discussed in Sec. \ref{sec:results} B.
%\section{The analytical form of the Berry connection} \label{app:approx}
%We choose the eigenstates in the gauge
%\begin{eqnarray}
%	u_{\pm}=
%	\frac{1}{2\epsilon}
%	\begin{pmatrix}
%	\pm({f_x}-i {f_y})\\
%	 2\epsilon\mp f_z
%	\end{pmatrix},
%\end{eqnarray}
%where $\epsilon=\sqrt{f_x^2+f_y^2+f_z^2}$, with the corresponding eigenenergy $E_{\pm}=f_0\pm\epsilon$. 
%Near the gap-closing point, the approximate expression for the Berry connection  $r_{12}^x=\frac{v_{12}^x}{i\omega_{12}}$ is
%\begin{eqnarray}
	%r_{11}^x-r_{22}^x &\approx& \frac{f_z}{2\epsilon^3}F(f_x,f_y),\\
%	r_{12}^x&\approx&-\frac{f_z}{4\epsilon^3}F(f_x,f_y)\label{eq:approx}
%\end{eqnarray}
%where $F(f_x,f_y)=+f_x\left[\sin(k_x)+\sin(k_x/2)\cos(\sqrt{3}y/2)\right]+f_y\left[\cos k_x  -\cos(k_x/2)\cos(\sqrt{3}y/2)\right]$. Across the band closure, only $f_z$ changes sign and lead to the abrupt change of the phase of $r_{12}^x$ from $0$ to $\pi$, leading to the jump in $\overline{I^{x,xx}_{12}}$ in Fig. \ref{fig:shdep}. 
\bibliography{references,referencesInpaper}      

\end{document}